\documentclass[%
 reprint,
superscriptaddress,
 amsmath,amssymb,
 aps,
]{revtex4-1}

\usepackage{graphicx}
\usepackage{dcolumn}
\usepackage{bm}
\usepackage{setspace}
\usepackage{xcolor}
\usepackage{soul}

\begin{document}

\preprint{APS/123-QED}

\title{Ultranarrow linewidth photonic-atomic laser}

\author{Wei Zhang}
\altaffiliation{These authors contributed equally to this work}
\affiliation{National Institute of Standards and Technology, 325 Broadway, Boulder, Colorado 80305, USA \\}
        
\author{Liron Stern}
\altaffiliation{These authors contributed equally to this work}
\affiliation{National Institute of Standards and Technology, 325 Broadway, Boulder, Colorado 80305, USA \\}
        
\author{David Carlson}
    \affiliation{National Institute of Standards and Technology, 325 Broadway, Boulder, Colorado 80305, USA \\}
\author{Douglas Bopp}
		\affiliation{National Institute of Standards and Technology, 325 Broadway, Boulder, Colorado 80305, USA \\}
\author{Zachary Newman}
		\affiliation{National Institute of Standards and Technology, 325 Broadway, Boulder, Colorado 80305, USA \\}
\author{Songbai Kang}
		\affiliation{National Institute of Standards and Technology, 325 Broadway, Boulder, Colorado 80305, USA \\}
\author{John Kitching}
		\affiliation{National Institute of Standards and Technology, 325 Broadway, Boulder, Colorado 80305, USA \\}
\author{Scott B. Papp}
        \email{scott.papp@colorado.edu}
        \affiliation{National Institute of Standards and Technology, 325 Broadway, Boulder, Colorado 80305, USA \\}
        \affiliation{Department of Physics, University of Colorado, Boulder, Colorado 80309, USA \\}

\date{\today}

\begin{abstract}
Lasers with high spectral purity can enable a diverse application space, including precision spectroscopy, coherent high-speed communications, physical sensing, and manipulation of quantum systems. Already, meticulous design and construction of bench Fabry-Perot cavities has made possible dramatic achievements in active laser-linewidth reduction, predominantly for optical-atomic clocks. Yet there is increasing demand for miniaturized laser systems operating with high performance in ambient environments. Here, we report a compact and robust photonic-atomic laser comprising a 2.5 cm long, 20,000 finesse, monolithic Fabry-Perot cavity integrated with a micromachined rubidium vapor cell. By leveraging the short-time frequency stability of the cavity and the long-time frequency stability of atoms, we realize an ultranarrow-linewidth laser that enables integration for extended measurements. Specifically, our laser supports a fractional-frequency stability of $1\times 10^{-13}$ at an averaging time of 20 ms, $7 \times 10^{-13}$ at 300 s, an integrated linewidth of 25 Hz that results from thermal noise, a Lorentzian linewidth as low as 0.06 Hz$^2$/Hz, and a passive vibration immunity as low as $10^{-10}$/g. Our work explores hybrid laser systems with monolithic photonic and atomic packages based on physical design.   
\end{abstract}

\maketitle

\section{Introduction}

Stable lasers that provide high-spectral purity are key components in a variety of advanced research direction, such as optical atomic clocks~\cite{Bloom2014,Hinkley2013}, gravitational wave detectors~\cite{Rana2014}, photonic microwave synthesizers~\cite{Fortier2011} and communication ~\cite{AlTaiyOL2014}. Innovation in these application areas can rely on laser phase-noise properties, which correspond to the spectral linewidth in the frequency domain ~\cite{Hall_1992}. To achieve a narrow integrated linewidth, state-of-the-art stable lasers are usually disciplined to high-finesse, Fabry-Perot cavities, and current performance demonstrations reach a linewidth remarkably well below 1 Hz ~\cite{HafnerOL2015,Matei2017,ZhangPRL2017,Robinson19Optica}. Achieving such performance requires a detailed design in order to isolate the laser from ambient fluctuations, including temperature, pressure, and vibration. Consequently, stable laser systems are comprised of relatively large size, weight and power consumption components, and are relatively complex to implement as they require vacuum, vibration isolation, and multiple layers of thermal isolation. In recent years there has been a significant effort to introduce portable stable lasers systems ~\cite{LeibrandtOE2011,ArgenceOE2012,SwieradSR2016,DavilaRodriguezOL2017} aiming for applications in which robustness and integration are primary considerations. However, it can still be a challenge to use these systems out of the laboratory environment due to their complexity. Realizing a compact laser source with high spectral purity and yet sufficiently simple and cost effective design is an active research field. Indeed, the reduction of the core dimensions of high performance stable lasers to the millimeter and centimeter scale have been demonstrated by several advanced techniques, such as whispering-gallery-mode resonator~\cite{Matsko2007,SavchenkovJOSAB07,AlnisPRA2011,Lim2017, Zhang2019}, chip-scale  resonator~\cite{LeeNC2013,SternOL2017}, semiconductor laser with injection locking~\cite{LiangNC2015}, and stimulated Brillouin scattering laser ~\cite{Lee2012,LohOE2016,LohOptica2019,Gundavarapu2019}

\begin{figure*}
\centering
\includegraphics[scale=.56]{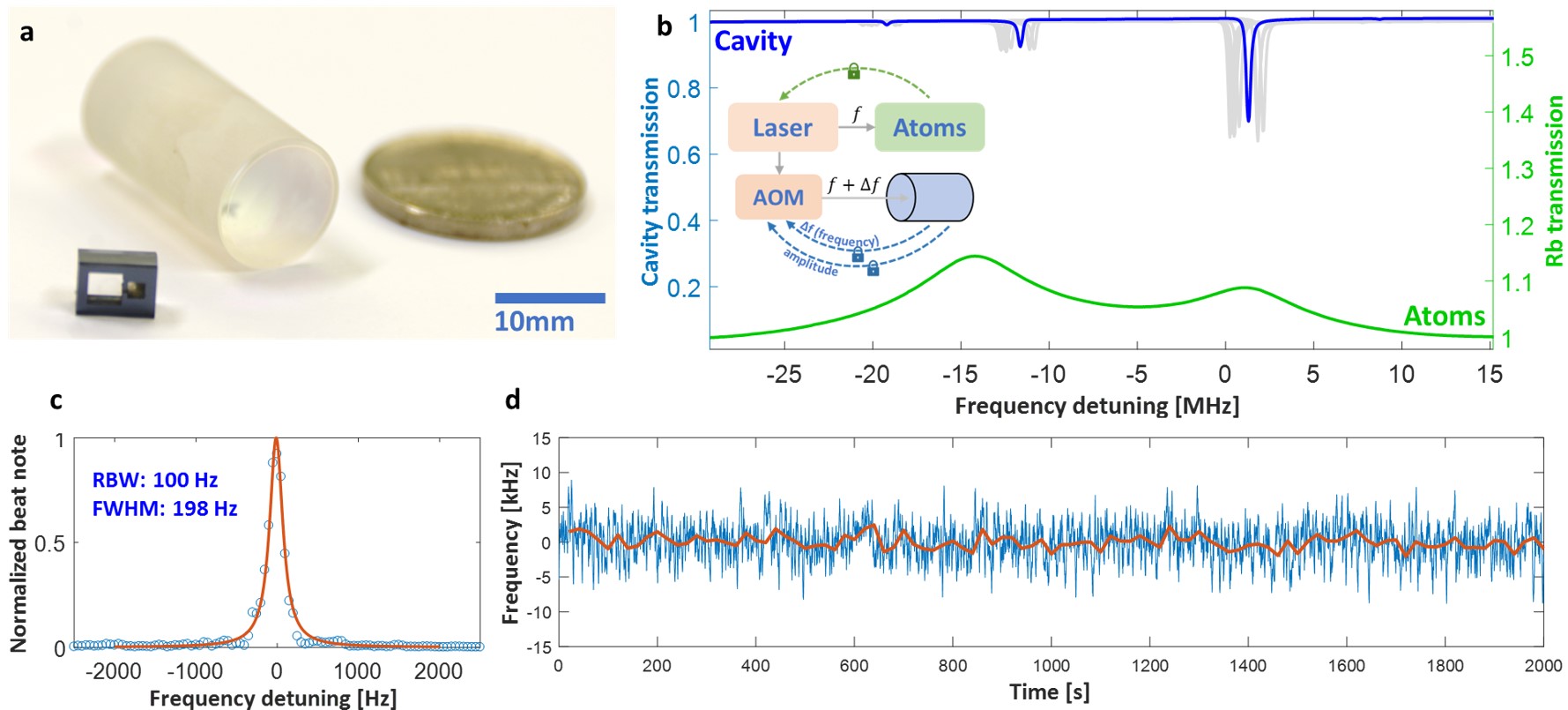}
\caption{Photonic-atomic ultranarrow-linewidth laser. a) Photograph of a fused-silica photonic cavity and a micromachined atomic vapor cell. b) Spectroscopy of the photonic cavity (blue curve) and the Doppler-free hyperfine transition of Rb (green curve). The frequency traces have been superimposed, using the atomic traces as an absolute reference, and they illustrate the thermo-optic frequency wander (gray traces) of the photonic-cavity and the difference in linewidth. Note that the horizontal axis is referred to a wavelength of 1560 nm. c) Instantaneous lineshape of the photonic-atomic laser, which is measured by an optical heterodyne with a stabilized EOM frequency comb. d) Measured frequency of the photonic-atomic laser as function of time showing a standard deviation of $\approx$ 2.5 kHz at a 20 ms averaging time (blue) and a standard deviation of 1 kHz for 10-s averages of the data (red)}
\end{figure*}

In this Article, we report a photonic-atomic laser, which is composed of a high-finesse, bulk cavity and a microfabricated Rb vapor cell. Our laser attains high coherence by stabilization to both the cavity resonance and the atomic resonance, hence hybridizing the photonic and atomic properties. The cavity is based on a bulk, fused-silica cylinder, and we apply superpolishing and low-loss reflection coatings on both facets to realize a finesse of $\approx20,000$. The absence of an evacuated center bore, which is essential in traditional cavity systems, simplifies our system by eliminating the need for vacuum. Vibration-immune design is mandatory to reduce vibration-induced frequency fluctuations. We achieve this goal by a novel design of the cavity holding structure for a net vibration sensitivity of $\approx10^{-10}$/$g$, where $g$ is earth's gravity. With these features we achieve a cavity-stabilized laser linewidth of 25 Hz in an ambient environment, which is dominated by thermorefractive noise of the cavity mode volume. The cavity resonance is also susceptible to temperature-induced frequency drift. The unvarying atomic transitions of Rb atoms with 6 MHz radiative lifetime ~\cite{Steck2010} cannot provide the same level of frequency discrimination as our cavity, yet Rb is an ideal system to correct medium and long timescale cavity drift. Therefore, we stabilize the cavity resonance, using all-optical thermal feedback derived from saturated-absorption Rb spectroscopy. Operating both system sections can require performance tradeoffs; the photonic-atomic laser linewidth is 200 Hz but the Allan deviation is $7\times10^{-13}$ in 300 s averaging periods. Due to the compact size of the bulk cavity and the micro-machined Rb atomic cell, our photonic-atomic laser offers a stable, accurate, and narrow linewidth source.




\section{overview of components and concept}
Figure 1 presents an overview of our photonic and atomic components and the main results of our experiments. A photograph of the fused-silica cavity alongside the micro-machined Rb atomic cell is presented in Fig. 1a. In experiments, we install these two components in aluminum enclosures. We operate the photonic-atomic laser by stabilizing a laser to the cavity, and referencing the cavity resonance with respect to a 780-nm atomic transition of Rb. 

In Fig. 1b, we present the spectroscopic response of the cavity and atomic cell versus optical frequency. The ion beam sputtered cavity coating reflectivity is designed and measured to be 0.99998. The scattering and loss due to the silica material is measured to be 130 part-per-million (ppm), and the cavity offers a linewidth of 200 kHz, a finesse of 20,000, and quality factor of $\approx1$ billion. Here we show the resonance modes (blue trace) in the reflection spectrum of the cavity. Due to the birefringence of the cavity, there are two polarization states for cavity fundamental modes, however they are well-resolved with a splitting of $\approx$ 12 MHz which is much larger than the cavity linewidth. We also show saturation spectroscopy of the Rb cell (green trace), corresponding to the $^{85}$Rb F=2 to F$'=1/2$ hyperfine transitions. To understand and portray the long-term evolution of the cavity and atomic Rb resonances, we acquire these traces many times over the course of a few hours using the atomic traces as an absolute reference. Hence, frequency drift appears to smear-out the sharp cavity resonances as shown by the gray traces while the relatively broad Rb atomic resonances with 6 MHz natural linewidth maintain a constant absolute optical frequency. Leveraging the advantages of these sub-systems is the central concept of our work, which enables a stable, ultranarrow-linewidth photonic-atomic laser.

We run the photonic-atomic laser by the procedure in the Fig. 1b inset. First, we lock a free-running semiconductor laser to the atomic resonance signal by direct feedback to laser frequency. Therefore a long-term average of the laser frequency yields the atomic-resonance frequency. Second, for convenience, we use an acousto-optic modulator (AOM) as an actuator to linewidth-narrow the semiconductor laser with respect to the cavity. In the final step, we discipline the cavity to the Rb atoms by slowly adjusting the intracavity power, which induces a thermo-optic frequency shift.

With an operating photonic-atomic laser, we measure its optical lineshape (Fig. 1c) and chart its output frequency (Fig. 1d) over the course of 30 minutes. At any suitable instant in time during this measurement the full-width half-maximum (FWHM) of the laser's optical lineshape is approximately 200 Hz; see Fig 1c. The time-series measurement of the laser frequency indicates a 2.5 kHz standard deviation of fluctuations (Fig. 1d blue trace) that can be attained in only 20 ms of averaging time. The red trace in Fig. 1d shows that the laser frequency is more stable by using 10 s averaging intervals, which filters out high-frequency noise. Calibration or complying with the International Bureau of Weights and Measures operational conditions for this atomic transition may provide improved accuracy. These results explain the short- and long-time stability of our system, offering narrow linewidth and kHz-level long-time frequency stability.

\section{Experimental Methods}

Figure 2a shows a schematic of the photonic-atomic laser and the measurement system that we use to characterize it. We measure the frequency stability of the cavity-stabilized laser by forming an optical heterodyne beat with a $\sim$ 1-Hz-linewidth reference continuous-wave (CW) laser whose frequency drifts on average at $\approx$ 0.1 Hz/s ~\cite{Baynes15}. We record and analyze the heterodyne beatnote both in the frequency domain with an FFT analyzer to determine the power-spectral density of frequency noise and in the time-domain with a frequency counter to determine the Allan deviation.

\begin{figure}[h!]
\centering
\includegraphics[width=\linewidth]{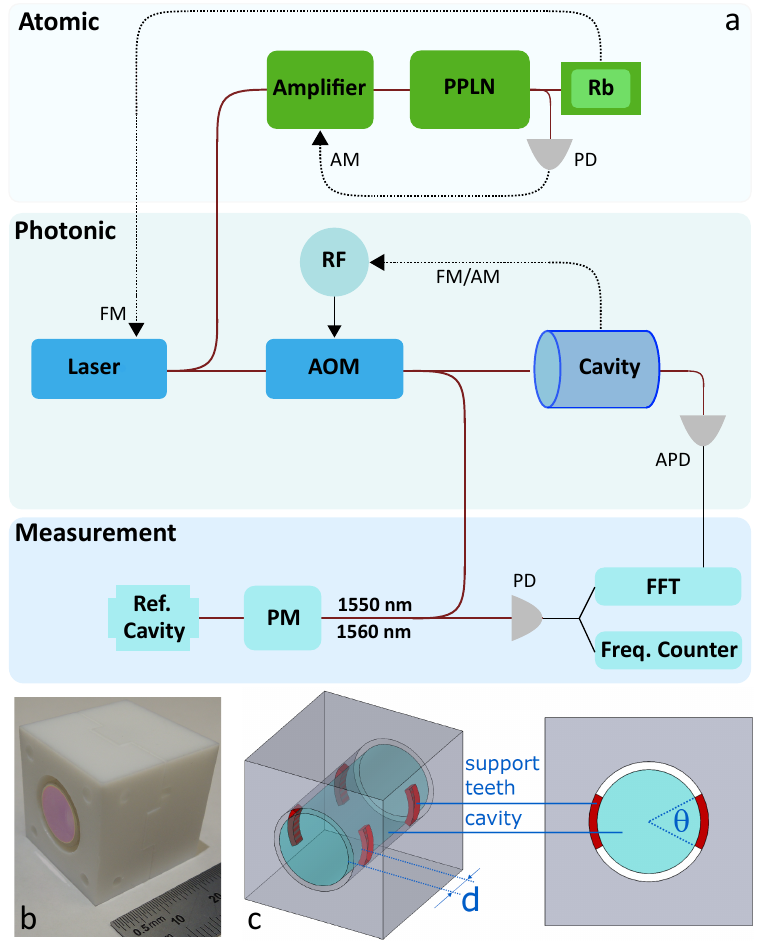}
\caption{Experimental scheme. a) A portion of the continuous-wave (CW) laser is amplified, frequency-doubled by a periodically poled lithium niobate (PPLN) device, and stabilized to the Rb Doppler-free D2 line using a current dither lock. An additional power stabilization servo is implemented using a photodetector (PD) and the current of the amplifier. The other part of CW laser is frequency-stabilized to the cavity by a PDH lock in which an acousto-optic modulator (AOM) is the frequency actuator. By controlling the AOM RF driving power the optical power injected to the cavity is controlled and via the thermo-optic effect modifies the cavity resonance frequency. The laser intensity noise is measured by an avalanche photodetector (APD). The laser frequency stability is characterized by optical heterodyne with a Hz-level reference laser. The beatnote is recorded by a fast Fourier transform (FFT) and a frequency counter. b) A photo of the cavity installed in the support made by Teflon. c) The transparent and the front view of the cavity-support to show the four support teeth (red). “d” is the distance between the cavity surface and the support tooth. The tooth is symmetrical relative to the midplane of the cavity and  “$\theta$” is the open angle of the tooth.}
\label{fig:2}
\end{figure}

In the atomic section, 1 mW of a CW laser at 1560 nm is amplified to 100 mW and then frequency-doubled to 780 nm by a periodically-poled lithium niobite (PPLN). The output power at 780 nm is $\approx$1 mW. By means of velocity selection in a counter propagating saturation spectroscopy scheme, we detect Doppler-free absorption lines with nearly lifetime-limited linewidth. We achieve such lineshapes by detecting a retroreflected beam that transverses a micro-machined mm-scale natural Rb vapor cell ~\cite{Liew2004}. The cell temperature is stabilized to approximately $60^\circ$ C, using coaxial heaters to reduce magnetic fields. Measurements of the temperature stability of the cell reveal a temperature deviation of $\pm 0.01 ^\circ$ C over the course of a few minutes  as the transitions in this implementation have a temperature coefficient of 40 kHz$/^\circ$C, such temperature stability implies a sub kHz stabilization limit. Light-shift-induced frequency shifts (measured to have a coefficient of $\approx$ 10 kHz/$\mu$W) are controlled in our system. 10\% of the 780 nm light before the vapor cell is detected and used to stabilize 780-nm laser power by actuating the current of the amplifier with a 10 Hz bandwidth. To frequency-lock the laser to Doppler free transition, the current of the CW laser is dithered at 10 kHz with an instantaneous frequency excursion of 400 kHz. By demodulating the retroreflected signal be means of a lock-in-amplifier, we lock the laser to the atomic transition with a feedback bandwidth of $\approx$ 500 Hz.


In the photonic section, the CW laser at 1560 nm is frequency-locked to the reference cavity with the Pound-Drever-Hall (PDH) locking scheme~\cite{Drever1983}. A fiber-packaged, waveguide electro-optic modulator (not shown in Fig. 2a) provides phase modulation at 20.5 MHz. An acousto-optic modulator (AOM) is the frequency actuator of which 10\% power is used as output of the photonic-atomic system. The rest of the power of the AOM output, after an inline isolator, is collimated and free-space coupled to the cavity. In experiment, the input power for the cavity is $\approx$ 200 $\mu$W. The reflection of the cavity is received by an avalanche photodetector (APD, not shown in Fig. 2a) and demodulated to generate the PDH error signal by which the frequency modulation port of the radio frequency (RF) synthesizer for AOM is driven for frequency lock. The feedback bandwidth is 200 kHz, which is rapid enough to substantially reduce the free-running CW laser-frequency noise. The transmission of the cavity is received by an APD to measure the laser intensity noise.


We install the photonic cavity in an hermetic aluminum enclosure (7 cm $\times$ 6 cm $\times$ 6 cm) without active or passive vibration isolation. We design the cavity support structure to minimize vibration-induced frequency noise, since the relatively small volume of the cavity itself does not enable such an optimization. The support structure is made of teflon, and it has the geometry shown in Fig. 2b. The transparent view of Fig. 2c highlights the four support teeth that directly contact the cavity. The overall vibration sensitivity of the cavity resonance frequency is optimized by adjusting $d$, the distance between the support teeth and the facet of the cavity, and $\theta$, the open angle of the tooth. According to the finite element simulation, we choose d=4.3 mm and $\theta$=55 degree. To measure the vibration sensitivity, the cavity with support structure is placed on an active isolation table which is driven by sinusoidal accelerations. An accelerometer calibrates the motion and the heterodyne beatnote is recorded for frequency response. With respect to earth's gravity $g$, we measure a vibration sensitivity of $1.7 \times 10^{-10}$/g along the gravitational direction and $1.4 \times 10^{-10}$/g in the horizontal plane, respectively. These sensitivities are consistent with our simulations and typically observed in optimized vacuum-gap Fabry-Perot cavities.

Another important consideration in our cavity support structure design is thermal isolation from room-temperature fluctuations, which induce drift in the cavity resonance frequency. We have not yet employed a multi-layer thermal isolator. Instead we utilize the low thermal conductivity of teflon, and minimize the direct contact area, each tooth is 2 mm in width, between teflon and the cavity. The cavity and its support form a thermal low-pass filter of which the measured time constant is $\approx$ 1100 s. As the temperature of the enclosure is actively controlled at $\approx$ milliKelvin level, the temperature-induced frequency fluctuation is expected to be $\approx$ $1.5 \times 10^{-12}$ $\tau$, where $\tau$ is the averaging time. 

By measuring the cavity-stabilized laser with respect to the atomic transition, we reveal a slow frequency drift of the cavity. Consequently, the output of the cavity servo-loop with the drifting DC component is sent to another servo-loop integrator which drives the amplitude modulation port of the RF synthesizer. We modulate the incident power coupled to the cavity to actuate its resonance frequency through the thermo-optic effect with a coefficient off 200 kHz/$\mu$W at the wavelength of 1.5 $\mu$m. Thus, aligning the cavity resonance to the Rb atomic transition is achieved by an all-optical scheme.

\section{result and discussion}

\begin{figure}[h!]
\centering
\includegraphics[width=\linewidth]{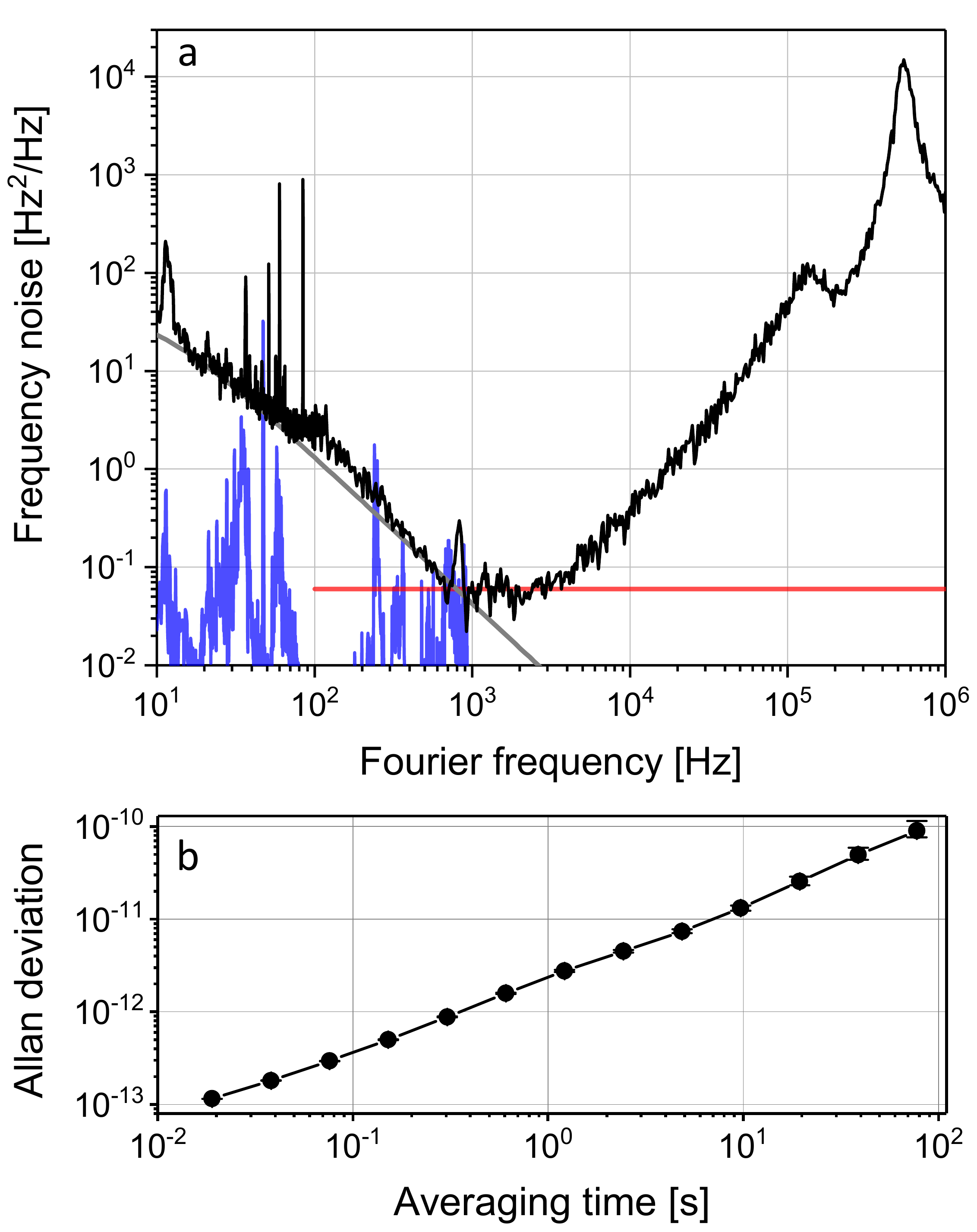}
\caption{Frequency stability characterization of the cavity-stabilized laser. a) Power-spectral density of laser frequency noise (black), detection-noise floor of the photodetector in PDH locking (red), inferred vibration-induced frequency noise (blue) and the cavity thermorefractive-noise floor (gray). b) Fractional-frequency stability measurement in time domain.}
\label{fig:3}
\end{figure}

In Figures 3 and 4, we characterize the optical-frequency stability of the cavity-stabilized laser and the photonic-atomic laser, respectively. To investigate the cavity, we disengage the phase modulation shown in Fig. 2a and frequency lock the CW laser to the cavity, using 10 $\mu$W of incident laser power. In Fig. 3a, the frequency-noise power spectral density of  the cavity-stabilized laser (black line) is 30 $\text{Hz}^\text {2}/ \text Hz$ at 10 Hz Fourier frequency, corresponding to an improvement by $10^{6}$ comparing to the free-running laser, and the Fourier-frequency dependence from 10 Hz to 1 kHz is consistent with what we predict for the cavity thermorefractive-noise spectrum (gray line) ~\cite{Matsko2007,Zhang2019}. The detection noise of the photodetector in the PDH locking (red line) limits the frequency noise at 0.06 $\text{Hz}^\text {2}/ \text Hz$ from 1 kHz to 3 kHz, which corresponds to a Lorentzian linewidth of only 0.2 Hz. Beyond a Fourier frequency of 3 kHz, the frequency noise is dominated by the inloop error of the 600 kHz bandwidth PDH lock. Due to the optimization of the cavity-vibration sensitivity, our estimate of the vibration-induced frequency noise (blue line) calculated by summing the vertical and horizontal directions in quadrature is well below the measured thermal-noise-limited frequency noise (black line).

We determine the laser linewidth to be 25 Hz by integrating the phase noise from infinity to the Fourier frequency corresponding to 1 $\text {rad}^\text {2}$. Therefore the integrated linewidth is limited by fundamental thermal noise. Future experiments searching for higher performance may need to contend with this thermal-noise limit in innovative ways, since the optical mode volume of such cavities that primarily sets the thermal-noise linewidth cannot be substantially improved.

Figure 3b presents fractional-frequency stability of the cavity-stabilized laser. We obtain the data for Allan deviation calculation by frequency counting the optical heterodyne beatnote with a dead-time free counter in so-called $\Lambda$ mode. In these measurements, we have not applied any correction for the linear frequency drift that is relatively constant over time periods that extend beyond the maximum range of averaging times shown here. The cavity-stabilized laser stability reaches  $1 \times 10^{-13}$ at $\tau$=20 ms and follows $\approx$ $1.5 \times 10^{-12}$ $\tau$ as predicted by the simulated temperature-induced frequency drift for $\tau>6\,\text{s}$.

The photonic cavity enables a narrow-linewidth laser, however temperature-induced frequency drift degrades the long-term frequency stability. Therefore, we stabilize the cavity resonance to a Rb transition which provides laser stability over periods of a few seconds and beyond. As mentioned earlier, this is achieved by controlling the power launched into the cavity, using an AOM. To achieve a stable and long-term lock, there are two aspects taken into account when choosing the laser power incident on the cavity. The first is the frequency drift rate of the cavity that varies from 100 Hz/s to 1000 Hz/s. We require sufficient control range through the incident power to maintain the photonic-atomic and cavity servo locks. Conversely, the laser-intensity-induced frequency noise is elevated due to the rising of the laser power. In our experiments, an overall compromise leads to an incident power of 200 $\mu$W.

\begin{figure}[h!]
\centering
\includegraphics[width=\linewidth]{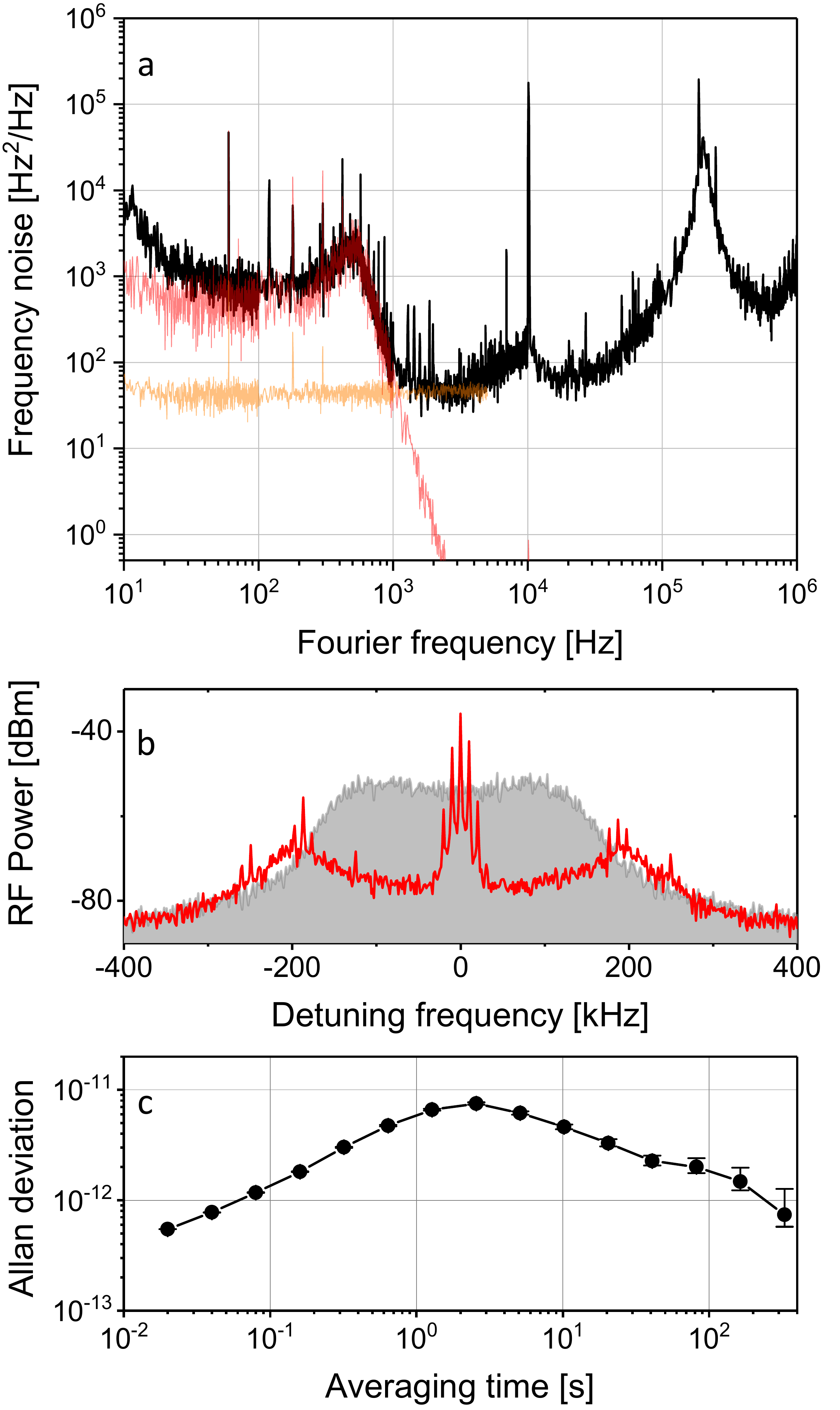}
\caption{Spectral coherence characterization of the photonic-atomic laser. a) Frequency noise (black), inloop error of the atomic servo (red), and laser-intensity induced noise (orange). b) Optical lineshape of the laser with dither lock (gray shaded area) and anti-dither by AOM. c) Fractional frequency stability measurement in time domain.}
\label{fig:4}
\end{figure}

Figure 4 shows our characterization of the photonic-atomic laser, using the locking scheme described in the Fig. 1b inset. In Fig. 4a, the frequency noise (black line) exhibits a dominant contribution for Fourier frequencies below 1 kHz. This is also evident by the inloop error (red line) of the servo stabilizing the laser to the Rb transition at 780 nm. Optimization of the modulation frequency, modulation depth and choice of detector may allow additional reduction of the frequency noise. These Fourier components increase the noise to $1000$ Hz$^2$/Hz. As a result, the laser linewidth defined by integrating the phase noise up to 1 $\text{rad}^\text{2}$ is 990 Hz. Coupling a larger optical power level to the cavity sets the laser-intensity-induced frequency noise (orange line) to a floor of 50 $\text{Hz}^\text {2}/ \text Hz$ up to 5 kHz. We infer this contribution from measurements of laser-intensity noise in the cavity transmission and the thermo-optic coefficient. The latter is found to be 200 kHz/$\mu$W due to the absorption of the built-in laser power circulating within the cavity, which in turn induces heat and thermo-optically shifts the cavity resonance. Beyond a Fourier frequency of 5 kHz, the inloop error of the PDH servo-loop locking the photonic cavity dominates the frequency-noise spectrum, including the servo bump at 200 kHz. 


Figure 4b shows the optical lineshape of the photonic-atomic laser. The laser current dither that we apply for stabilization to a Rb transition at 780 nm broadens the spectral width to as much as 400 kHz; see the gray shaded area in Fig. 4b. By implementing the cavity lock at the same time as the atom lock, the laser linewidth is narrowed substantially to 200 Hz; see the red trace in Fig. 4b. The residual peaks in the red trace correspond to the 10 kHz dither signal. 

Figure 4c shows the fractional frequency stability that our photonic-atomic laser achieves. This Allan deviation provides the clearest summary of our hybrid laser system, capturing the crossover between passive stability of the cavity and intrinsic stability of the atomic transition. In comparison to our results without atomic stabilization (Fig. 3b), the stability is degraded for $20\, \textrm{ms}<\tau<1 \,\text{s}$. However, for $\tau>1 \,\text{s}$, the stability is well below $10^{-11}$ with a slope of $\tau^{-1/2}$ showing the impact of the atomic lock. For example, the frequency stability at $\tau$=100 s is improved by two orders of magnitude compared to the result shown in Fig. 3b.

\section{conclusion}
We have demonstrated a compact photonic-atomic laser, featuring high spectral purity and long-term frequency stability traceable to Rb atoms. The miniaturized components are relatively cost-effective, robust, and offer a  passive vibration immune design as low as $10^{-10}$/g. The cavity supports short-term stability reaching $1 \times 10^{-13}$ at $\tau$=20 ms, leading to 25 Hz integrated linewidth. Interfacing such a cavity with atoms, our system mostly preserves the short-time stability of the cavity and shows a stability of  $7 \times 10^{-13}$ at $\tau$=300 s, and an instantaneous linewidth of 200 Hz. 

Our results indicate a path for achieving even better performance. It is noteworthy that the finesse of the monolithic Fabry-Perot cavity is now limited by the material optical loss, which we measure to be 130 ppm. Since fused silica with optical losses $<$10 ppm should be available, it is conceivable to substantially increase the cavity $Q$ to above 10 billion. Cavity frequency drift associated with ambient temperature fluctuations may also be further suppressed by a more aggressive thermal low-pass filter. Moreover, the atomic-Rb system could also be operated at higher performance and with absolute frequency calibration with control of systematic effects. Indeed, fractional stability as low as $4 \times 10^{-13}\tau^{-1/2}$ have been reported ~\cite{TPA2018Burke}, using a two-photon Rb transition at 778 nm with a cm-scale cell. Recently, such transitions have been demonstrated in micromachined cells supporting stabilities as low as $4 \times 10^{-12}\tau^{-1/2}$ ~\cite{Newman2019}. Thus, optimizing the cavity and the implemented atomic line, provides a path to a thermal-noise-limited, low drift, and accurate laser with Hz-level integral linewidth.


\section*{Funding Information}
This work is supported by AFOSR under award number FA9550-16-1-0016, the DARPA ACES program, and NIST.  

\section*{Acknowledgments}
The authors thank Vincent Maurice and Su-Peng Yu for comments on the paper. The authors thank Vincent Maurice for help in software and technical support.



\bibliography{reference}

\end{document}